\documentclass[twocolumn,pra,showpacs,showkeys]{revtex4}
\usepackage{amssymb}
\usepackage{amsfonts}
\usepackage{amsmath}
\usepackage{graphicx}

\begin{document}

\title{Bipartite entanglement induced by classically-constrained quantum
dissipative dynamics}
\author{Adri\'{a}n A. Budini}
\affiliation{Consejo Nacional de Investigaciones Cient\'{\i}ficas y T\'{e}cnicas, Centro
At\'{o}mico Bariloche, Avenida E. Bustillo Km 9.5, (8400) Bariloche,
Argentina}
\date{\today }

\begin{abstract}
The properties of some complex many body systems can be modeled by
introducing in the dissipative dynamics of each single component a set of
kinetic constraints that depend on the state of the neighbor systems. Here,
we characterize this kind of dynamics for two quantum systems whose
independent dissipative evolutions are defined by a Lindblad equation. The
constraints are introduced through a set of projectors that restrict the
action of each single dissipative Lindblad channel to the state of the other
system. Conditions that guaranty a classical interpretation of the kinetic
constraints are found. The generation and evolution of entanglement is
studied for two optical qubits systems. Classically constrained dissipation
leads to a stationary state whose degree of entanglement depends on the
initial state. Nevertheless, independently of the initial conditions, a
maximal entangled state is generated when both systems are subjected to the
action of local Hamiltonian fields that do not commutate with the
constraints. The underlying physical mechanism is analyzed in detail.
\end{abstract}

\maketitle



\section{Introduction}

Superposition of different states corresponding to a bipartite quantum
system may lead to entanglement, that is, quantum states whose statistical
properties can not be reproduced with local stochastic variables. Due to its
central role in quantum information and quantum computation \cite{nielsen},
in the last years and ever-increasing interest has been paid to\ its
characterization \cite{horo}.

Superposition of bipartite quantum states are degraded by interaction with
uncontrollable degrees of freedom \cite{breuerbook}. In contrast to quantum
decoherence, the decay of entanglement, measured for example by concurrence 
\cite{wooters}, has non-usual properties such as its finite time decay \cite%
{yu}\ as well as the possibility of its sudden appearance at posterior times 
\cite{ficekrevival}. Non-standard decay properties were also found in the
classical \cite{vedral} and quantum \cite{discord} contributions of the
total correlation (mutual information) between two systems \cite%
{clasica,maniscalco}.

Even when the environment destroy the nonlocal properties of entanglement,
it may also play a constructive role. The creation of entanglement by the
action of a common bath was studied in many different physical situations 
\cite%
{pianiMarkov,jako,lendi,walter,knight,braun,wang,3level,ayunet,marzolino,jakobis,tanas,ficek,xu,obada}
. Special interest have been paid to optical systems such as the Dicke model 
\cite{jakobis,tanas,ficek} where the dissipative decay dynamics is induced
by interaction with the background electromagnetic field.

Added to the individual irreversible dynamics, dissipation of systems
embedded in complex many body arranges may involves extra elements. 
For example, in quantum aggregates \cite{eisfeld} the effects of a bath on a
subsystem can be dependent on the state of another site \cite{thilagam}.
This kind of constrained dissipation also arises in glassy systems. In fact,
in contrast to static 
disordered interactions, 
glassiness can also be induced by dynamical constraints 
\cite{andersen,pietro,david,chandler,ritort}. In a classical context, this
situation is usually modeled by a Markovian master equation where the
transition rates of each system depend on the state of the neighbors systems 
\cite{ritort}. In Ref. \cite{beatriz} it was introduced a quantum version
where the underlying evolution is given by a Lindblad equation and a set of
projection operators introduce the constraints.

Constrained dissipation lead to an effective interaction between subsystems.
As the dissipative dynamics admits a Markovian description \cite%
{ritort,beatriz}, we expect that some entanglement may be generated by this
mechanism \cite{pianiMarkov}. Motivated by the previous physical situations,
in this paper we study the production and evolution of entanglement for
systems whose dissipative evolution is a constrained one. Our main
theoretical goal is to characterize which kind of stationary entangled
states may be generated by the constraints when they admit a classical
interpretation \cite{ritort}. The classicality property seems to be a very
restrictive condition for the generation of entanglement. Nevertheless,
focusing our analysis on a bipartite dynamics, we find that the joint action
of classical dissipative constraints and local unitary evolutions may drive
the dynamics to maximal entangled states. 

We introduce a generalized definition of constrained dissipation, where the
dynamical action of each system's dissipative channel can only happens when
the other system is in a given quantum state \cite{beatriz}. Conditions that
guarantee the \textquotedblleft classicality\textquotedblright\ of the
kinetic constraints is provided. In such a case, the system's reduced
evolutions are defined by a Lindblad rate equation \cite{rate,SMS}. The
entanglement generation is analyzed for two optical-like qubits systems
whose individual decay dynamics can only happens when the other system is in
the lower state. The entanglement is characterized both in the transitory
and stationary regimes. In order to provide an deeper characterization of
the dynamics, added to the entanglement behavior, we also study its relation
with the quantum and classical correlations build up between both systems 
\cite{luiz,angeles,tavis,chen,wu}.

The paper is organized as follows. In Sect. II we provide a generalized
definition of constrained dissipation for bipartite dynamics. Conditions
that guarantee the classicality of the kinetic constraints is provided. In
Sec. III we analyze the entanglement evolution for two optical-like qubits.
In Sec. IV, the interplay between the constraints an local external fields
that lead to a maximal entangled state is investigated. In Sec. V we provide
the Conclusions. In the appendix we briefly review the utilized entanglement
measure as well as the definitions of quantum and classical correlations.

\section{Constrained dissipation}

We consider two systems $A$ and $B,$ both of them coupled to independent
Markovian reservoirs. The dissipative Lindblad evolution \cite{breuerbook}
induced by each bath is denoted by $\mathcal{L}_{A}[\rho ]$ and $\mathcal{L}%
_{B}[\rho ].$ The bipartite density matrix describing both systems is $\rho
_{AB}(t).$ In order to focus on the dissipative structure, we assume that
both Lindblad superoperators commutate with their respective system unitary
evolutions. Hence, in an interaction representation, the evolution of $\rho
_{AB}(t)$ can be written as%
\begin{equation}
\frac{d\rho _{AB}(t)}{dt}=\mathcal{L}_{A}[\rho _{AB}(t)]+\mathcal{L}%
_{B}[\rho _{AB}(t)].  \label{Lindblad}
\end{equation}%
Each contribution is defined by the expressions 
\begin{subequations}
\label{Lin}
\begin{eqnarray}
\mathcal{L}_{A}[\rho ] &=&\frac{1}{2}\sum_{i}\gamma _{A}^{i}([A_{i},\rho
A_{i}^{\dag }]+[A_{i}\rho ,A_{i}^{\dag }]), \\
\mathcal{L}_{B}[\rho ] &=&\frac{1}{2}\sum_{i}\gamma _{B}^{i}([B_{i},\rho
B_{i}^{\dag }]+[B_{i}\rho ,B_{i}^{\dag }]),
\end{eqnarray}%
where $\{\gamma _{A}^{i},\gamma _{B}^{i}\}$\ are the dissipative rates of
each dissipative channel defined by the operators $\{A_{i},B_{i}\}.$ In each
sum, the index run from one up to the dimension of the system Hilbert space.
For simplifying the notation, from now on we assume $\dim (\mathcal{H}%
_{A})=\dim (\mathcal{H}_{B}).$

The operators $\{A_{i}\}$ and $\{B_{i}\}$ only act, respectively, on the
Hilbert space of system $A$ and $B.$ Hence, the bipartite evolution Eq. (\ref%
{Lindblad}) is defined under the association 
\end{subequations}
\begin{equation}
A_{i}\rightarrow A_{i}\otimes \mathrm{I}_{B},\ \ \ \ \ \ \ B_{i}\rightarrow 
\mathrm{I}_{A}\otimes B_{i}.  \label{Independent}
\end{equation}%
The identity operators $(\mathrm{I}_{A},\mathrm{I}_{B})$ indicates the
absence of any correlation between the dynamics induced by each bath.

Constrained dissipation means that the action of each dissipative channel
(operators $A_{i}$ and $B_{i})$ only happen when the other system ($B$ and $A
$) is in a given subspace or quantum state. These kinetic constraints are
introduced through the replacements%
\begin{equation}
A_{i}\rightarrow A_{i}\otimes \mathcal{Q}_{i},\ \ \ \ \ \ \ B_{i}\rightarrow 
\mathcal{P}_{i}\otimes B_{i},  \label{Constraint}
\end{equation}%
where $\{\mathcal{P}_{i}\}$ and $\{\mathcal{Q}_{i}\}$ are orthogonal
projectors operators, $\mathcal{P}_{i}\mathcal{P}_{j}=\delta _{ij}\mathcal{P}%
_{i},$ $\mathcal{Q}_{i}\mathcal{Q}_{j}=\delta _{ij}\mathcal{Q}_{i},$ acting
on systems $A$ and $B$ respectively. With these definitions it become
evident that the action of a given reservoir over its system is dynamically
restricted or conditioned by the state of the other system. Notice that in
contrast with the approach of Ref. \cite{beatriz}, here each Lindblad
channel may be characterized by a different projector. Trivially, the
present generalization recover that case by taking $\mathcal{P}%
_{i}\rightarrow \mathcal{P},$ and $\mathcal{Q}_{i}\rightarrow \mathcal{Q.}$

Some general characterization of the constrained dissipation induced by the
operators (\ref{Constraint}) can be achieved after assuming some properties
for the set of projectors operators.

\subsection{Stationary properties}

We assume that the un-constrained evolution, Eqs. (\ref{Lin}) and (\ref%
{Independent}), has a unique and separable stationary state%
\begin{equation}
\rho _{AB}^{\infty }=\lim_{t\rightarrow \infty }\rho _{AB}(t)=\rho
_{A}^{\infty }\otimes \rho _{B}^{\infty }.
\end{equation}%
Hence, we may ask about the effects of the constraints (\ref{Constraint}) on
this state. While a general answer is not possible, under the conditions%
\begin{equation}
\lbrack \mathcal{P}_{i},\rho _{A}^{\infty }]=[\mathcal{Q}_{i},\rho
_{B}^{\infty }]=0,  \label{Commutation}
\end{equation}%
it is simple to realize that the stationary state remains the same.
Nevertheless, it is not possible to guarantee that any initial condition
relax to the same stationary state. In general, only a subset of initial
conditions $\rho _{AB}(0)\in \{\mathcal{H}_{0}\}$ fulfill this condition,%
\begin{equation}
\rho _{AB}^{\infty }|_{\{\mathcal{H}_{0}\}}=\lim_{t\rightarrow \infty }\rho
_{AB}(t)|_{\{\mathcal{H}_{0}\}}=\rho _{A}^{\infty }\otimes \rho _{B}^{\infty
},
\end{equation}%
where $\mathcal{H}_{0}\in \mathcal{H}_{AB}$ is a subspace over the set of
all possible initial conditions for $A$ and $B.$ This property arises
because the dynamic becomes reducible, that is, the Hilbert space decompose
into different subsets between which no transition is possible. Free
decoherence subspaces as well as dark states may characterize the subspace
complementary to $\mathcal{H}_{0}.$ These features are\ explicitly shown in
the next sections.

\subsection{Classical constraints}

Given the quantum nature of both systems, the constraints can only be
understood when acting in Hilbert space. Here, we found which conditions
permit us to read the kinetic constraints in a classical way \cite{ritort}.

Let assume the completeness properties%
\begin{equation}
\sum_{i}\mathcal{P}_{i}=\mathrm{I}_{A},\ \ \ \ \ \ \ \sum_{i}\mathcal{Q}_{i}=%
\mathrm{I}_{B}.  \label{projector}
\end{equation}%
Therefore, the partial matrixes $\rho _{A}(t)=\mathrm{Tr}_{B}[\rho _{AB}(t)]$
and $\rho _{B}(t)=\mathrm{Tr}_{A}[\rho _{AB}(t)]$ can be written as 
\begin{subequations}
\label{parciales}
\begin{eqnarray}
\rho _{A}(t) &=&\sum_{i}\mathrm{Tr}_{B}[\mathcal{Q}_{i}\rho _{AB}(t)]\equiv
\sum_{i}\rho _{A}^{(i)}(t), \\
\rho _{B}(t) &=&\sum_{i}\mathrm{Tr}_{A}[\mathcal{P}_{i}\rho _{AB}(t)]\equiv
\sum_{i}\rho _{B}^{(i)}(t).
\end{eqnarray}%
The matrix $\rho _{A}^{(i)}(t)$ $[\rho _{B}^{(i)}(t)]$ can be read as the
conditional evolution of system $A$ $(B)$ given that system $B$ $(A)$ is in
the state defined by the projector $\mathcal{Q}_{i}$ $(\mathcal{P}_{i}).$ In
general, the evolution of these conditional states can not be written in a
closed way without involving other matrix elements (coherences) contained in 
$\rho _{AB}(t).$ A classical interpretation of the constraints can only be
achieved when these last objects does not participate in the dynamics of the
auxiliary states. That is, it should be possible to write a closed evolution
for the set $\{\rho _{A}^{(i)}(t)\}$ as well as for the set $\{\rho
_{B}^{(i)}(t)\}.$

From Eqs. (\ref{Lin}) and (\ref{Constraint}), it is simple to realize that
the previous classicality condition is satisfied when the projectors are
closed under the transitions induced by each dissipative channel, 
\end{subequations}
\begin{equation}
A_{j}^{\dag }\mathcal{P}_{i}A_{j}=\sum_{k}\alpha _{ik}^{j}\mathcal{P}_{k},\
\ \ \ \ \ \ \ B_{j}^{\dag }\mathcal{Q}_{i}B_{j}=\sum_{k}\beta _{ik}^{j}%
\mathcal{Q}_{k}.  \label{closure}
\end{equation}%
Here, $\alpha _{ik}^{j}$ and $\beta _{ik}^{j}$ are real positive
coefficients that also satisfy $\alpha _{ii}^{j}=\beta _{ii}^{j}=0.$ These
relations, jointly with Eq. (\ref{projector}), implies that $A_{j}^{\dag
}A_{j}=\sum_{i,k}\alpha _{ik}^{j}\mathcal{P}_{k},$ and $B_{j}^{\dag
}B_{j}=\sum_{i,k}\beta _{ik}^{j}\mathcal{Q}_{k}.$ By introducing the closure
condition (\ref{closure}) in the constrained evolution, Eqs. (\ref{Lindblad}%
) and (\ref{Constraint}), we get $[\rho _{A}^{(i)}(t)\rightarrow \rho
_{A}^{(i)}]$ 
\begin{subequations}
\label{LindbladRates}
\begin{eqnarray}
\frac{d\rho _{A}^{(i)}}{dt} &=&\gamma _{A}^{i}\left( A_{i}\rho
_{A}^{(i)}A_{i}^{\dag }-\frac{1}{2}\{A_{i}^{\dag }A_{i},\rho
_{A}^{(i)}\}_{+}\right)  \label{RateA} \\
&&+\sum_{j,k}b_{ij}^{k}\mathcal{P}_{k}\rho _{A}^{(j)}\mathcal{P}_{k}-\frac{1%
}{2}\sum_{j,k}b_{ji}^{k}\{\mathcal{P}_{k},\rho _{A}^{(i)}\}_{+},  \notag
\end{eqnarray}%
jointly with the symmetrical expression $[\rho _{B}^{(i)}(t)\rightarrow \rho
_{B}^{(i)}]$%
\begin{eqnarray}
\frac{d\rho _{B}^{(i)}}{dt} &=&\gamma _{B}^{i}\left( B_{i}\rho
_{B}^{(i)}B_{i}^{\dag }-\frac{1}{2}\{B_{i}^{\dag }B_{i},\rho
_{B}^{(i)}\}_{+}\right)  \label{RateB} \\
&&+\sum_{j,k}a_{ij}^{k}\mathcal{Q}_{k}\rho _{B}^{(j)}\mathcal{Q}_{k}-\frac{1%
}{2}\sum_{j,k}a_{ji}^{k}\{\mathcal{Q}_{k},\rho _{B}^{(i)}\}_{+}.  \notag
\end{eqnarray}%
With $\{\cdots \}_{+}$ we denote an anticommutation operation. Furthermore,
the rate coefficients read 
\end{subequations}
\begin{equation}
a_{ij}^{k}=\gamma _{A}^{k}\alpha _{ij}^{k},\ \ \ \ \ \ \ \ \
b_{ij}^{k}=\gamma _{B}^{k}\beta _{ij}^{k}.
\end{equation}

Equations (\ref{RateA}) and (\ref{RateB}) demonstrate that under the
conditions (\ref{closure}) the time evolution of the reduced dynamic of each
system can be written in a closed way without involving in an explicit way
matrix elements of the other system. These dynamical equations are a
particular case of Lindblad rate equations \cite{rate}, which describe the
more general evolution of a quantum system coupled to a set of classical
degrees of freedom \cite{SMS}. Here, the underlying classical feature of
these evolutions can explicitly be shown by introducing the expectation
values%
\begin{equation}
p_{j}^{i}(t)\equiv \mathrm{Tr}_{A}[\rho _{A}^{(i)}(t)\mathcal{P}_{j}],\ \ \
\ \ \ \ q_{j}^{i}(t)\equiv \mathrm{Tr}_{B}[\rho _{B}^{(i)}(t)\mathcal{Q}%
_{j}].
\end{equation}%
By using the completeness conditions (\ref{projector}) and the definitions (%
\ref{parciales}) it follow the normalizations $\sum_{ij}p_{j}^{i}(t)=1$ and $%
\sum_{ij}q_{j}^{i}(t)=1.$ Furthermore, Eq. (\ref{parciales}) implies the
relation $q_{j}^{i}(t)=p_{i}^{j}(t).$ From the Lindblad rate equations (\ref%
{LindbladRates}) we get the \textit{classical rate equation} $%
[p_{j}^{i}(t)\rightarrow p_{j}^{i}]$%
\begin{equation}
\frac{dp_{j}^{i}}{dt}=\sum_{k}(a_{jk}^{i}p_{k}^{i}-a_{kj}^{i}p_{j}^{i})+%
\sum_{k}(b_{ik}^{j}p_{j}^{k}-b_{ki}^{j}p_{j}^{i}).\ \ \ \ \ \ 
\label{maestraClasica}
\end{equation}%
The evolution of $p_{j}^{i}$ $(q_{j}^{i})$ consists of two contributions.
The first one is induced by the diagonal contributions of Eq. (\ref{RateA})
[Eq.(\ref{RateB})] leading to the classical transitions $k\rightarrow j$
with rates $a_{jk}^{i}$ $(b_{jk}^{i}).$ These contributions take into
account the fluctuation of system $A$ $(B)$ induced by its own reservoir.
The second contribution follows from the nondiagonal term of Eq. (\ref{RateA}%
) [Eq.(\ref{RateB})], being defined by the rates $b_{ik}^{j}$ $(a_{ik}^{j}),$
corresponding to the transitions $k\rightarrow i.$ These terms take into
account the fluctuations of the constraints of system $A$\ $(B)$ induced by
the dissipative dynamics of system $B$ $(A).$

We conclude that the quantum dynamical constraints Eq. (\ref{Constraint})
are classical if their influence can read from a (transition) rate
evolution, here defined by Eq. (\ref{maestraClasica}). In the next section,
we study the generation and evolution of entanglement for two qubits
characterized by this kind of classically constrained dissipation.

\section{Classical constrained dissipation in two qubits systems}

We consider to qubit systems $A$ and $B.$ The state of each one is
represented in the basis $\left\vert \pm \right\rangle .$ Their density
matrix $\rho _{AB}(t)$ evolve as 
\begin{equation}
\frac{d\rho _{AB}(t)}{dt}=\frac{-i}{\hbar }[H,\rho _{AB}(t)]+\mathcal{L}%
[\rho _{AB}(t)].  \label{evolution}
\end{equation}%
In the basis $\left\vert \pm \right\rangle ,$ and in absence of any external
field, the unitary contribution is defined by the Hamiltonian%
\begin{equation}
H=\frac{\hbar \omega _{A}}{2}\sigma _{z}\otimes \mathrm{I}_{B}+\frac{\hbar
\omega _{B}}{2}\mathrm{I}_{A}\otimes \sigma _{z},  \label{isolated}
\end{equation}%
where $\sigma _{z}$ is the z-Pauli matrix and $\omega _{s}$ $(s=A,B)$ are
the transition frequencies. A unique dissipative channel define the Lindblad
contribution of each system. Hence, $\mathcal{L}[\rho ]$ is written as%
\begin{equation}
\mathcal{L}[\rho ]=\frac{1}{2}\sum_{s=A,B}\gamma _{s}([V_{s},\rho
V_{s}^{\dag }]+[V_{s}\rho ,V_{s}^{\dag }]),  \label{LindbladTwo}
\end{equation}%
where $\gamma _{s}$\ is the dissipative rate of each system. The operators $%
\{V_{s}\}$ characterize the interaction of each system with the environment.
For optical arranges, they must be taken as \cite{breuerbook}%
\begin{equation}
V_{A}=\sigma \otimes \mathrm{I}_{B},\ \ \ \ \ \ V_{B}=\mathrm{I}_{A}\otimes
\sigma ,
\end{equation}%
where $\sigma =\left\vert -\right\rangle \left\langle +\right\vert $ is the
lowering operator. In this case, independently of the initial condition, the
stationary state is%
\begin{equation}
\rho _{AB}^{\infty }=\lim_{t\rightarrow \infty }\rho _{AB}(t)=\left\vert
--\right\rangle \left\langle --\right\vert ,  \label{downdown}
\end{equation}%
that is, independently of the initial conditions, both systems end in the
lower state.

\subsection{Classical constraints}

The dynamical constraints are written as%
\begin{equation}
V_{A}=\sigma \otimes \mathcal{P}_{B},\ \ \ \ \ \ \ V_{B}=\mathcal{P}%
_{A}\otimes \sigma .  \label{projectores}
\end{equation}%
We consider the symmetric projectors case%
\begin{equation}
\mathcal{P}_{A}=\left\vert -\right\rangle \left\langle -\right\vert ,\ \ \ \
\ \ \ \ \ \ \ \mathcal{P}_{B}=\left\vert -\right\rangle \left\langle
-\right\vert .  \label{constraints}
\end{equation}%
With these constraints, the decay dynamics $(\left\vert +\right\rangle
\rightsquigarrow \left\vert -\right\rangle )$ of each system is only
possible when the other system is in the lower state. We remark that these
operators and projectors can be read as a two qubit version of the
Fredickson-Andersen model (with periodic boundary conditions) or East model
of Ref. \cite{beatriz} [see Eqs. (4) and (5) of that paper]. Nevertheless,
due to the optical motivation, here the effective temperature is cero.
Furthermore, the projectors, instead of the upper state, here are defined by
the lower state. In spite of the states (lower or upper), the classicality
condition (\ref{closure}) is satisfied by Eqs. (\ref{projectores}) and (\ref%
{constraints}). A quantum constraint should to violates condition (\ref%
{closure}), such as for example taking both projectors as $\mathcal{P}%
=\left\vert x_{\pm }\right\rangle \left\langle x_{\pm }\right\vert ,$ where $%
\left\vert x_{\pm }\right\rangle =(1/\sqrt{2})(\left\vert +\right\rangle \pm
\left\vert -\right\rangle )$ are the eigenstates of the $x$-Pauli matrix. In
this case, all results of section II do no apply.

Given that the classicality condition (\ref{closure}) is satisfied by Eqs. (%
\ref{projectores}) and (\ref{constraints}), the evolution of the partial
projected matrixes must be given by a Lindblad rate equation. The partial
matrixes $\rho _{A}(t)=\mathrm{Tr}_{B}[\rho _{AB}(t)]$ and $\rho _{B}(t)=%
\mathrm{Tr}_{A}[\rho _{AB}(t)]$ can be written as%
\begin{equation}
\rho _{A}(t)=\rho _{A}^{-}(t)+\rho _{A}^{+}(t),\ \ \ \ \ \ \ \ \rho
_{B}=\rho _{B}^{-}(t)+\rho _{B}^{+}(t),
\end{equation}%
where $\rho _{A}^{-}(t)=\mathrm{Tr}_{B}[\mathcal{P}_{B}\rho _{AB}(t)],$ $%
\rho _{A}^{+}(t)=\mathrm{Tr}_{B}[(\mathrm{I}_{B}-\mathcal{P}_{B})\rho
_{AB}(t)],$ $\rho _{B}^{-}(t)=\mathrm{Tr}_{A}[\mathcal{P}_{A}\rho _{AB}(t)],$
and $\rho _{B}^{+}(t)=\mathrm{Tr}_{A}[(\mathrm{I}_{A}-\mathcal{P}_{A})\rho
_{AB}(t)].$ From Eqs. (\ref{LindbladTwo}) and (\ref{projectores}), in an
interaction representation with respect to the Hamiltonian (\ref{isolated}),
we consistently get 
\begin{subequations}
\label{ParalelaPerpen}
\begin{eqnarray}
\frac{d\rho _{A}^{-}(t)}{dt} &=&\gamma _{A}\mathrm{L}[\rho
_{A}^{-}(t)]+\gamma _{B}\mathcal{P}_{A}\rho _{A}^{+}(t)\mathcal{P}_{A}, \\
\frac{d\rho _{A}^{+}(t)}{dt} &=&-\frac{1}{2}\gamma _{B}\{\mathcal{P}%
_{A},\rho _{A}^{+}(t)\}_{+},
\end{eqnarray}%
where $\mathrm{L}[\rho ]\equiv (1/2)([\sigma ,\rho \sigma ^{\dag }]+[\sigma
\rho ,\sigma ^{\dag }]).$ Symmetrically,%
\begin{eqnarray}
\frac{d\rho _{B}^{-}(t)}{dt} &=&\gamma _{B}\mathrm{L}[\rho
_{B}^{-}(t)]+\gamma _{A}\mathcal{P}_{B}\rho _{B}^{+}(t)\mathcal{P}_{B}, \\
\frac{d\rho _{B}^{+}(t)}{dt} &=&-\frac{1}{2}\gamma _{A}\{\mathcal{P}%
_{B},\rho _{B}^{+}(t)\}_{+}.
\end{eqnarray}%
Given the bipartite initial condition $\rho _{AB}(0),$ these equations
completely define the reduced dynamics of each system. Correlations between
them can only be characterized from the bipartite Lindblad equation (\ref%
{evolution}).

By denoting the product basis as 
\end{subequations}
\begin{subequations}
\label{Puros}
\begin{eqnarray}
\left\vert 1\right\rangle &=&\left\vert ++\right\rangle ,\ \ \ \ \ \ \
\left\vert 3\right\rangle =\left\vert -+\right\rangle , \\
\left\vert 2\right\rangle &=&\left\vert +-\right\rangle ,\ \ \ \ \ \ \
\left\vert 4\right\rangle =\left\vert --\right\rangle .
\end{eqnarray}%
the dissipative evolution [Eqs. (\ref{LindbladTwo}) and (\ref{projectores})]
reads 
\end{subequations}
\begin{equation}
\mathcal{L}[\rho ]=-\left( 
\begin{array}{cccc}
0 & \frac{\gamma _{A}}{2}\rho _{12} & \frac{\gamma _{B}}{2}\rho _{13} & 0 \\ 
\frac{\gamma _{A}}{2}\rho _{21} & \gamma _{A}\rho _{22} & \gamma \rho _{23}
& \frac{\gamma _{A}}{2}\rho _{24} \\ 
\frac{\gamma _{B}}{2}\rho _{31} & \gamma \rho _{32} & \gamma _{B}\rho _{33}
& \frac{\gamma _{B}}{2}\rho _{34} \\ 
0 & \frac{\gamma _{A}}{2}\rho _{42} & \frac{\gamma _{B}}{2}\rho _{43} & 
-(\gamma _{A}\rho _{22}+\gamma _{B}\rho _{33})%
\end{array}%
\right) ,  \label{computacional}
\end{equation}%
where $\rho _{ij}=\left\langle i\right\vert \rho \left\vert j\right\rangle ,$
$i,j=1,2,3,4,$ and $\gamma \equiv (\gamma _{A}+\gamma _{B})/2.$ The
influence of the constraints can be easily understood by analyzing the
diagonal elements of this equation, which in turn define the classical
evolution Eq. (\ref{maestraClasica}). Evidently, the dynamics take place if
at least one of the systems is in the lower state. Therefore, the pure state 
$\left\vert ++\right\rangle \left\langle ++\right\vert $ does not decay at
all and is preserved by the evolution. Equivalently, it is a dark state. On
the other hand, in the complementary (Hilbert) space, the dynamic is
attracted toward the stationary state in absence of constraints, that is, $%
\left\vert --\right\rangle \left\langle --\right\vert .$ In fact, condition (%
\ref{Commutation}) is also satisfied in this case.

\subsection{Entanglement of the stationary state}

Given that the dissipative and unitary generators commutate, the stationary
state corresponding to the Lindblad evolution (\ref{evolution}) can
straightforwardly be obtained from Eq. (\ref{computacional}). For arbitrary
initial conditions, we get%
\begin{equation}
\rho _{AB}^{\infty }=\lim_{t\rightarrow \infty }\rho _{AB}(t)=\left( 
\begin{array}{cccc}
p & 0 & 0 & c \\ 
0 & 0 & 0 & 0 \\ 
0 & 0 & 0 & 0 \\ 
c^{\ast } & 0 & 0 & 1-p%
\end{array}%
\right) ,  \label{rhoInfi}
\end{equation}%
where the matrix elements reads 
\begin{subequations}
\begin{eqnarray}
p &=&\left\langle ++\right\vert \rho _{AB}(0)\left\vert ++\right\rangle , \\
c &=&\left\langle ++\right\vert \rho _{AB}(0)\left\vert --\right\rangle .
\end{eqnarray}%
The stationary partial matrixes are $\rho _{A}^{\infty }=\rho _{B}^{\infty
}=diag\{p,1-p\}$ which in agreement with Eqs. (\ref{ParalelaPerpen}), imply $%
\rho _{A}^{-}(\infty )=\rho _{B}^{-}(\infty )=(1-p)\left\vert -\right\rangle
\left\langle -\right\vert $ and $\rho _{A}^{+}(\infty )=\rho _{B}^{+}(\infty
)=p\left\vert +\right\rangle \left\langle +\right\vert .$

The stationary state\ $\rho _{AB}^{\infty }$\ depends on the initial
conditions only when the dark state $\left\vert ++\right\rangle $ is
populated at the initial time. In this situation, some entanglement may be
found in the long-time limit. Even more, from Eq. (\ref{rhoInfi}) we deduce
that two Bell states are preserved by the dynamics. That is, for $\left\vert
\Phi _{\pm }\right\rangle $ at any time (in an interaction representation)
the density matrix satisfies 
\end{subequations}
\begin{equation}
\rho _{AB}(t)=\rho _{AB}(0)=\left\vert \Phi _{\pm }\right\rangle
\left\langle \Phi _{\pm }\right\vert .  \label{Invariantes}
\end{equation}%
As usual, the Bell basis is denoted as $\left\vert \Phi _{\pm }\right\rangle
\equiv \frac{1}{\sqrt{2}}(\pm \left\vert ++\right\rangle +\left\vert
--\right\rangle ),$ and $\left\vert \Psi _{\pm }\right\rangle \equiv \frac{1%
}{\sqrt{2}}(\pm \left\vert +-\right\rangle +\left\vert -+\right\rangle ).$

Evidently, the states (\ref{Invariantes}) are preserved because they are a
superposition of the dark state and the stationary state corresponding to
the complementary Hilbert space. In general, the stationary state is not a
maximal entangled one. Its degree of entanglement can be measured by
concurrence $C[\rho _{AB}^{\infty }]$ \cite{wooters}. In Appendix we get 
\begin{equation}
C[\rho _{AB}^{\infty }]=2|c|.  \label{ConcuEstacion}
\end{equation}

The capacity of the dissipative dynamics for \textquotedblleft
generating\textquotedblright\ entanglement can be characterized by
maximizing the concurrence $C[\rho _{AB}^{\infty }]$ given that the systems
begin in an arbitrary separable state. Therefore, we take 
\begin{equation}
\rho _{AB}(0)=\rho _{\mathbf{n}_{a}}\otimes \rho _{\mathbf{n}_{b}},
\label{polaris}
\end{equation}%
where each matrix is defined from%
\begin{equation}
\rho _{\mathbf{n}}=\frac{1}{2}(\mathrm{I}+\lambda \sigma _{\mathbf{n}}).
\label{RhoEne}
\end{equation}%
Here, $\mathrm{I}$ is the 2x2 identity matrix, the parameter $\lambda \in
\lbrack -1,1]$ gives the degree of purity of $\rho _{\mathbf{n}},$ $\mathrm{%
Tr}[\rho _{\mathbf{n}}^{2}]=(1+\lambda ^{2})/2,$ and $\sigma _{\mathbf{n}}$
is the Pauli matrix corresponding to an arbitrary direction in the Bloch
sphere \cite{nielsen}, $\sigma _{\mathbf{n}}=\{\{\cos (\theta ),\sin (\theta
)e^{-i\phi }\},\{\sin (\theta )e^{+i\phi },-\cos (\theta )\}\},$ being
defined by the polar angles $\theta \in \lbrack 0,\pi ]$ and $\phi \in
\lbrack 0,2\pi ].$ From Eqs. (\ref{polaris}) and (\ref{RhoEne}), the
parameters of the stationary state (\ref{rhoInfi}) read 
\begin{subequations}
\begin{eqnarray}
p &=&\frac{1}{4}[1+\lambda _{a}\cos (\theta _{a})][1+\lambda _{b}\cos
(\theta _{b})], \\
c &=&\frac{1}{4}\lambda _{a}\lambda _{b}\exp [-i(\phi _{a}+\phi _{b})]\sin
(\theta _{a})\sin (\theta _{b}).
\end{eqnarray}%
Considering these expressions it is possible to find the set of values of $%
\{\lambda _{a},\lambda _{b},\theta _{a},\theta _{b},\phi _{a},\phi _{b}\}$
that maximize the entanglement of $\rho _{AB}^{\infty },$ that is $|c|$ [Eq.
(\ref{ConcuEstacion})]. We get $\lambda _{a}=\pm 1,$ $\lambda _{b}=\pm 1,$ $%
\theta _{a}=\theta _{b}=\pi /2,$\ and arbitrary values of $\phi _{a}$ and $%
\phi _{b},$ delivering 
\end{subequations}
\begin{equation}
\max_{\rho _{AB}(0)\,\mathrm{sep}}\{C[\rho _{AB}^{\infty }]\}=\frac{1}{2}.
\label{ConcuMax}
\end{equation}%
Hence, over the set of all separable initial conditions, pure states lying
on the $x-y$ plane lead to the maximal possible concurrence in the long time
limit.

\subsection{Evolution of entanglement and quantum-classical correlations}

In Fig. 1 we plot the evolution of the concurrence taking the separable
symmetric initial condition $\rho _{AB}(0)=\rho _{y}\otimes \rho _{y}.$ Each
state is defined by $\rho _{y}=(1/2)(\mathrm{I}+\lambda \sigma _{y}),$ where 
$\sigma _{y}$ is the y-Pauli matrix and $\lambda \in \lbrack -1,1].$ The
stationary state [Eq. (\ref{rhoInfi})] becomes characterized by the
coefficients $p=1/4$ and $c=-\lambda ^{2}/4.$ From Eq. (\ref{ConcuEstacion})
the stationary concurrence reads%
\begin{equation}
C[\rho _{AB}^{\infty }]=\frac{\lambda ^{2}}{2}.
\end{equation}%
Consistently with the previous analysis, $C[\rho _{AB}^{\infty }]=1/2$ only
when $\lambda =\pm 1,$ that is, when both systems begin in an eigenstate of $%
\sigma _{y}.$ Similarly to the Dicke model \cite{ficek,luiz}, Fig. 1 also
shows that for all mixed states $(\lambda \neq \pm 1)$ there exist a time
delay before entanglement emerges. The birth time, $\tau _{0},$ is related
to the degree of purity. When $\gamma _{A}=\gamma _{B}=\gamma ,$ we get%
\begin{equation}
\gamma \tau _{0}=2\log (1/|\lambda |).  \label{birth}
\end{equation}%
%
%
%
%
%
%
%
%
%
%
%
%
%
%
%
%
%
%
%
%
%
%
%
\begin{figure}[tbp]
\includegraphics[bb=0 0 450 270,angle=0,width=10cm]{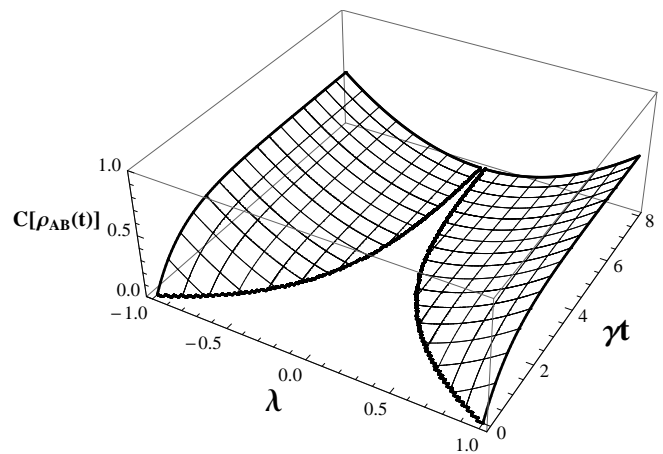}
\caption{Evolution of the concurrence $C[\protect\rho _{AB}(t)]$ as a
function of time for the constrained evolution defined by Eq. (\protect\ref%
{computacional}). The initial conditions are $\protect\rho _{AB}(0)=\protect%
\rho _{y}\otimes \protect\rho _{y}$ where $\protect\rho _{y}=(1/2)(\mathrm{I}%
+\protect\lambda \protect\sigma _{y}).$ The parameters are $\protect\gamma %
_{A}=\protect\gamma _{B}=\protect\gamma .$ The plot does not depends on the
transition frequencies $\protect\omega _{A}$ and $\protect\omega _{B}.$ }
\end{figure}
In order to understand this result, in Fig 2(a), for $\lambda =1/2,$ we plot
the four bipartite populations, while in Fig. 2(b) we plot the coherences
(in a interaction representation). For the chosen initial conditions, the
time evolution of the populations is the same of any value of $\lambda .$ In
both plots, consistently with Eq. (\ref{computacional}), the constant matrix
elements correspond to $\left\langle ++\right\vert \rho _{AB}(t)\left\vert
++\right\rangle ,$ and $\left\langle ++\right\vert \rho _{AB}(t)\left\vert
--\right\rangle .$ In Fig. 2(c) we plot the concurrence and also the
classical and quantum correlations (see Appendix). Comparing with the
previous plot [Fig. 2(b)], it becomes evident that, for the considered
initial conditions and parameters, the concurrence becomes not null when the
absolute value of coherence $\left\langle ++\right\vert \rho
_{AB}(t)\left\vert --\right\rangle =c$ [or equivalently $\left\langle
--\right\vert \rho _{AB}(t)\left\vert ++\right\rangle ]$ is larger than the
absolute value of any other coherence. This condition, from Eq. (\ref%
{computacional}), leads to the analytical result (\ref{birth}). 
\begin{figure}[tbp]
\includegraphics[bb=35 160 500
540,angle=0,width=9cm]{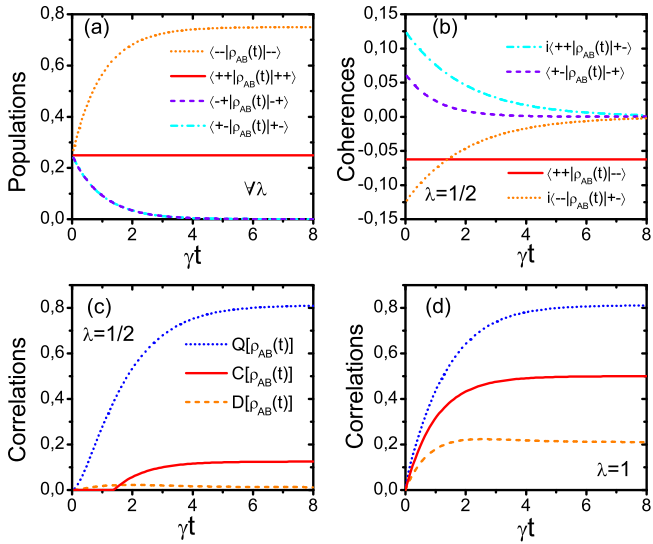}
\caption{Plot of matrix elements and correlations as a function of time
associated to the evolution Eq. (\protect\ref{computacional}). The initial
conditions and parameters are the same than in Fig. 1. (a) Populations,
which do not depend on the value of the purity parameter $\protect\lambda .$
(b) Coherences in a representation interaction for $\protect\lambda =1/2.$
(c) Concurrence $C[\protect\rho _{AB}(t)]$ (full line) classical $\mathcal{Q}%
[\protect\rho _{AB}(t)]$ (dotted line) and quantum $\mathcal{D}[\protect\rho %
_{AB}(t)]$ (dashed line) correlations for $\protect\lambda =1/2.$ (d) The
same objects for $\protect\lambda =1.$}
\end{figure}

Fig. 2(c) also shows that the belated appearance of entanglement is preceded
by the buildup of strong classical correlations and also moderate quantum
correlations. These properties are similar to that found in Ref. \cite{luiz}%
\ for an optical Dicke model. Nevertheless, here the quantum correlations
have a minor weight when compared to the classical ones. On the other hand,
in contrast to the Dicke model, the constrained dynamics is unable to create
any entanglement when the initial condition is any of the pure separable
states (\ref{Puros}).

In Fig. 2(d) we show the time dependence of the concurrence and correlations
for $\lambda =1,$ that is for a separable pure initial condition. In
agreement with Eq. (\ref{birth}) the entanglement delay time is null. On the
other hand, consistently with the previous comments, at shorter times the
classical correlation growing rate is larger than the quantum one. 

Most of results developed in this section also apply when the kinetic
constraints (\ref{projectores}) are defined by the projectors $\mathcal{P}%
_{A}=\mathcal{P}_{B}=\left\vert +\right\rangle \left\langle +\right\vert .\ $%
For example, Eq. (\ref{ConcuMax}) applies while Eq. (\ref{Invariantes})
remains valid after the replacement $\left\vert \Phi _{\pm }\right\rangle
\rightarrow \left\vert \Psi _{\pm }\right\rangle .$ Similarly, with these
projectors the dynamics is unable to create any entanglement if the initial
conditions are the separable pure states (\ref{Puros}).

\section{Interplay between external local excitation and classically
constrained dissipation}

In the previous section we showed that, even when the constraints (\ref%
{constraints})\ admits a classical interpretation [Eq. (\ref{ParalelaPerpen}%
)], the dynamics can generate some entanglement in the stationary regime.
Nevertheless, the stationary concurrence depends on the initial conditions.
In this section, we are interested in characterizing how these properties
are affected when the systems are coupled to external local Hamiltonian
fields that do not commutate with the corresponding projectors. In this
situation, the projector's average values can not be obtained from a
classical master equation like Eq. (\ref{maestraClasica}). On the other
hand, the external fields by they self cannot generate any entanglement.
Hence, the interplay between constrained dissipation and the external
excitation is the central ingredient to analyze.

\subsection{Maximal entangled stationary state}

Optical (fluorescent) two-level systems can be resonantly excited with an
external laser field, such that in an interaction representation their
Hamiltonian becomes $\hbar \Omega \sigma _{x}/2,$ where the Rabi frequency $%
\Omega $ measures the laser strength \cite{breuerbook}. Taking this kind of
external excitation for each system, the Hamiltonian [Eq. (\ref{isolated})]
of the optical-like qubits considered in the previous section becomes%
\begin{equation}
H=\frac{\hbar \Omega _{A}}{2}\sigma _{x}\otimes \mathrm{I}_{B}+\frac{\hbar
\Omega _{B}}{2}\mathrm{I}_{A}\otimes \sigma _{x},  \label{HRabi}
\end{equation}%
where $\sigma _{x}$ is the x-Pauli matrix, $\Omega _{A}$ and $\Omega _{B}$\
measure the interaction of each system with the external local fields. Eq. (%
\ref{HRabi}) is valid when\ both transition frequencies [Eq. (\ref{isolated}%
)] are the same, $\omega _{A}=\omega _{B}.$ At the end of this section we
analyze the consequences of raising up this and other symmetry conditions.
On the other hand, notice that the constraints Eq. (\ref{constraints}) do
not commutate with the local Hamiltonian (\ref{HRabi}).

Trivially, the eigenvectors of $H$ are defined by the external product of
the eigenvectors of $\sigma _{x},$ that is $\left\vert x_{\pm }\right\rangle
\otimes \left\vert x_{\pm }\right\rangle $ with eigenvalues $\pm \Omega =\pm
(\Omega _{A}+\Omega _{B})/2,$ and $\left\vert x_{\pm }\right\rangle \otimes
\left\vert x_{\mp }\right\rangle $ with eigenvalues $\pm \delta \Omega =\pm
(\Omega _{A}-\Omega _{B})/2.$ For symmetric local excitations%
\begin{equation}
\Omega _{A}=\Omega _{B}=\Omega ,  \label{Omega}
\end{equation}%
a degeneracy arises for the null eigenvalue, $\pm \delta \Omega =0.$ It is
simple to demonstrate that the Bell states $\left\vert \Phi
_{-}\right\rangle $ and $\left\vert \Psi _{-}\right\rangle $ lay in the
plane of degeneracy, that is, they are eigenvectors of the Hamiltonian with
null eigenvalue. Taking into account the invariance defined by Eq. (\ref%
{Invariantes}), it follows that when $\rho _{AB}(0)=\left\vert \Phi
_{-}\right\rangle \left\langle \Phi _{-}\right\vert ,$ the dynamics defined
by Eqs. (\ref{projectores}) and (\ref{HRabi}) satisfies $\rho _{AB}(t)=\rho
_{AB}(0).$ Nevertheless, for the present situation there not exist a dark
state. In fact, independently of the initial condition, under the joint
action of the local Hamiltonians and the constrained dissipation, the
stationary state is%
\begin{equation}
\rho _{AB}^{\infty }=\lim_{t\rightarrow \infty }\rho _{AB}(t)=\left\vert
\Phi _{-}\right\rangle \left\langle \Phi _{-}\right\vert ,
\label{EstacionBell}
\end{equation}%
that is, $\rho _{AB}^{\infty }$ is a maximally entangled pure Bell state.
The uniqueness of the stationary state can be proved in a pure mathematical
way (the density matrix evolution has a unique eigenoperator with null
eigenvalue). Below, we understand this fact from a dynamical point of view.

\subsection{Collective coherent-dissipative dynamics}

The interplay between the unitary [Eq. (\ref{HRabi})] and the dissipative
[Eq. (\ref{computacional})] dynamics that leads to the stationary state (\ref%
{EstacionBell}) can be analyzed in a simpler way in the collective basis 
\begin{subequations}
\label{BaseColectiva}
\begin{eqnarray}
\left\vert 1\right\rangle &=&\left\vert ++\right\rangle ,\ \ \ \ \ \ \
\left\vert 3\right\rangle =\left\vert \Psi _{-}\right\rangle , \\
\left\vert 2\right\rangle &=&\left\vert \Psi _{+}\right\rangle ,\ \ \ \ \ \
\ \left\vert 4\right\rangle =\left\vert --\right\rangle .
\end{eqnarray}%
In this basis, the Hamiltonian (\ref{HRabi}) reads 
\end{subequations}
\begin{equation}
H=\frac{1}{\sqrt{2}}\left( 
\begin{array}{cccc}
0 & \Omega & \delta \Omega & 0 \\ 
\Omega & 0 & 0 & \Omega \\ 
\delta \Omega & 0 & 0 & -\delta \Omega \\ 
0 & \Omega & -\delta \Omega & 0%
\end{array}%
\right) ,  \label{Hamilton}
\end{equation}%
where as before $\Omega =(\Omega _{A}+\Omega _{B})/2,$ and $\delta \Omega
=(\Omega _{A}-\Omega _{B})/2.$ Hence, when the condition (\ref{Omega}) is
meet $(\delta \Omega =0),$ the external field only couples (coherently) the
states $\left\vert ++\right\rangle \overset{\Omega }{\leftrightarrow }%
\left\vert \Psi _{+}\right\rangle $ and $\left\vert \Psi _{+}\right\rangle 
\overset{\Omega }{\leftrightarrow }\left\vert --\right\rangle .$ 
\begin{figure}[tbp]
\includegraphics[bb=55 450 455
800,angle=0,width=6.5cm]{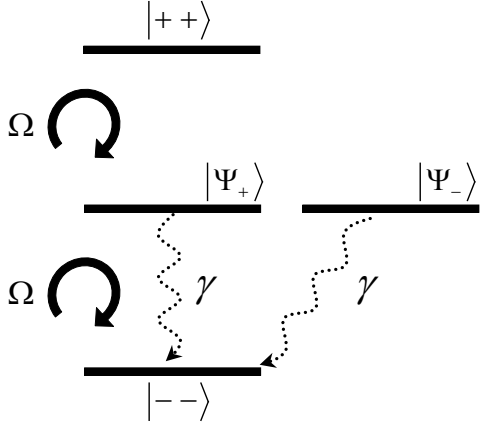}
\caption{Level scheme for the dynamics defined by Eqs. (\protect\ref%
{evolution}), (\protect\ref{projectores}), (\protect\ref{constraints}), and (%
\protect\ref{HRabi}). The dissipative and coherent couplings are valid under
the symmetry conditions $\Omega _{A}=\Omega _{B}=\Omega ,$ and $\protect%
\gamma _{A}=\protect\gamma _{B}=\protect\gamma $ (see text).}
\end{figure}

The interplay between the coherent and dissipative dynamics can be more
easily described by assuming $\gamma _{A}=\gamma _{B}=\gamma .$ While this
is not a necessary condition for the validity of Eq. (\ref{EstacionBell}),
under that condition the dissipative dynamics written in the collective
basis (\ref{BaseColectiva}) can be read from Eq. (\ref{computacional}) under
the replacements $\gamma _{A}\rightarrow \gamma ,$ $\gamma _{B}\rightarrow
\gamma .$ In Fig. 3 we show the levels scheme as well as the dissipative and
coherent coupling valid for $\gamma _{A}=\gamma _{B}=\gamma $ and $\Omega
_{A}=\Omega _{B}=\Omega .$ The state $\left\vert \Psi _{-}\right\rangle $ is
not affected by the unitary dynamic and it can only decay to the state $%
\left\vert --\right\rangle .$ Thus, the dynamics can be reduced to the
remaining three collective states. The dynamics of the states $\left\vert
\Psi _{+}\right\rangle \ $and $\left\vert --\right\rangle $ is equivalent to
a fluorescent system \cite{breuerbook} with Rabi frequency $\Omega $ and
natural decay $\gamma .$ The role of upper and lower states is played by $%
\left\vert \Psi _{+}\right\rangle $ and $\left\vert --\right\rangle $
respectively. As a consequence of the constraints, the state $\left\vert
++\right\rangle $ does not decay. Nevertheless, it is coupled coherently to
the state $\left\vert \Psi _{+}\right\rangle .$ $\left\vert ++\right\rangle $
and $\left\vert \Psi _{+}\right\rangle $ play the role of upper and lower
levels respectively. 
\begin{figure}[tbp]
\includegraphics[bb=55 30 420 555,angle=0,width=7cm]{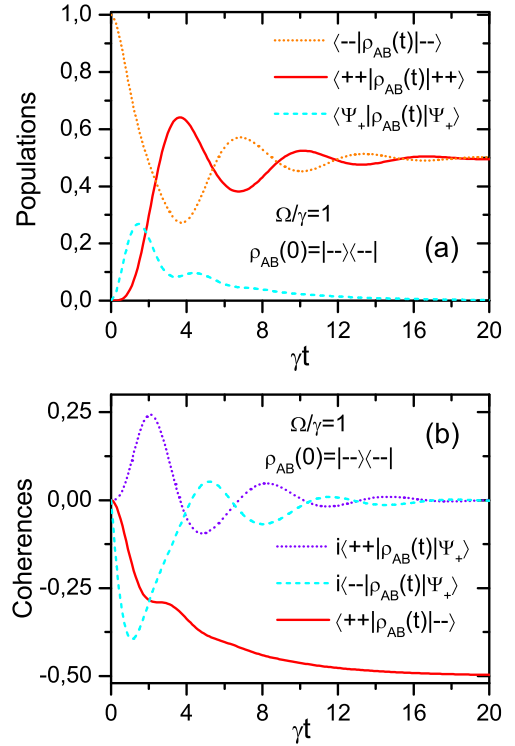}
\caption{(a) Collective populations as function of time. (b) Evolution of
collective coherences. These plots correspond to the dynamics (\protect\ref%
{evolution}) with the Hamiltonian Eq. (\protect\ref{HRabi}) and the
constraints defined by Eqs. (\protect\ref{projectores}) and (\protect\ref%
{constraints}). The initial condition is $\protect\rho _{AB}(0)=\left\vert
--\right\rangle \left\langle --\right\vert .$ The parameters are $\Omega
_{A}=\Omega _{B}=\Omega $ and $\protect\gamma _{A}=\protect\gamma _{B}=%
\protect\gamma ,$ with $\Omega /\protect\gamma =1.$}
\end{figure}

The opposite roles played by $\left\vert \Psi _{+}\right\rangle ,$ jointly
with uncoupling of the state $\left\vert ++\right\rangle $ from the
reservoir decay dynamics leads to (i) an asymptotic depopulation of the
state $\left\vert \Psi _{+}\right\rangle ,$ (ii) a raising of the $%
\left\vert ++\right\rangle \leftrightarrow \left\vert --\right\rangle $\
coherence, and (iii) an asymptotic vanishing of the difference of the $%
\left\vert ++\right\rangle $ and $\left\vert --\right\rangle $\ occupations
(populations). This coherent-dissipative mechanism drives the system to the
Bell state $\left\vert \Phi _{-}\right\rangle $ independently of the initial
condition, Eq. (\ref{EstacionBell}).

In order to clarify the previous mechanism, in Fig. 4 we plot the time
dependence of the collective populations and coherences. The density matrix
evolution (\ref{evolution}) is defined by the Hamiltonian Eq. (\ref{HRabi})
and the constraints introduced in Eqs. (\ref{projectores}) and (\ref%
{constraints}). The parameters are $\Omega _{A}=\Omega _{B}=\Omega $ and $%
\gamma _{A}=\gamma _{B}=\gamma ,$ with $\Omega /\gamma =1.$ At the initial
time both systems are in their respective lower states, $\rho
_{AB}(0)=\left\vert --\right\rangle \left\langle --\right\vert .$ Notice
that without the external fields the systems remain at all times in this
state. Furthermore, with this initial condition and parameters the state $%
\left\vert \Psi _{-}\right\rangle $ is never populated.

Fig. 4(a) shows the dissipative-coherent interplay that lead to a vanishing
of the $\left\vert \Psi _{+}\right\rangle $ population. Furthermore, as
expected the population of states $\left\vert ++\right\rangle $\ and $%
\left\vert --\right\rangle $\ oscillates with opposite phases, while in the
long time regime both of them become equal to one half. In Fig. 4(b), we
show the collective coherences. Coherences involving the state $\left\vert
\Psi _{+}\right\rangle $ vanish asymptotically, while the coherence between
the states $\left\vert ++\right\rangle \leftrightarrow \left\vert
--\right\rangle $ converges to minus one half. The asymptotic value of the
coherence $\left\langle ++\right\vert \rho _{AB}(t)\left\vert
--\right\rangle $ and the populations $\left\langle ++\right\vert \rho
_{AB}(t)\left\vert ++\right\rangle ,$ $\left\langle --\right\vert \rho
_{AB}(t)\left\vert --\right\rangle ,$ indicate that the bipartite dynamics
achieve the pure Bell state $\left\vert \Phi _{-}\right\rangle .$ 
\begin{figure}[tbp]
\includegraphics[bb=65 30 420 550,angle=0,width=7cm]{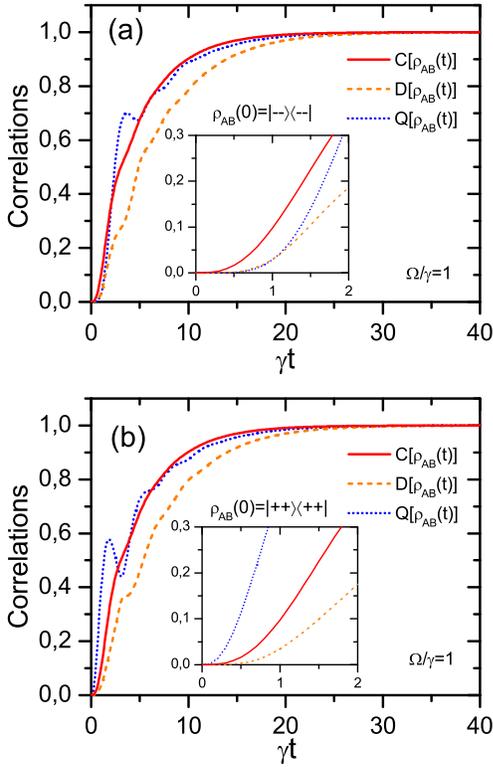}
\caption{Plot of Concurrence $C[\protect\rho _{AB}(t)]$ (full line)
classical $\mathcal{Q}[\protect\rho _{AB}(t)]$ (dotted line) and quantum $%
\mathcal{D}[\protect\rho _{AB}(t)]$ (dashed line) correlations. The
parameters and evolution of $\protect\rho _{AB}(t)$ are the same than in
Fig. 4. (a) Time dependence corresponding to the initial condition $\protect%
\rho _{AB}(0)=\left\vert --\right\rangle \left\langle --\right\vert .$ (b)
Initial condition $\protect\rho _{AB}(0)=\left\vert ++\right\rangle
\left\langle ++\right\vert .$ In both insets we show the short time behavior
of the correlations.}
\end{figure}

In Fig. 5(a), we show the time evolution of concurrence and the
classical-quantum correlations corresponding to the dynamic shown in Fig. 4,
that is, for the initial condition $\rho _{AB}(0)=\left\vert --\right\rangle
\left\langle --\right\vert .$ As the dynamics in the long time regime
reaches a maximal entangled state, all correlations converge to one. On the
other hand, the short time behavior strongly departs from that shown in Fig.
2(d). In fact, here the concurrence $C[\rho _{AB}(t)]$ grows much faster
than the classical and quantum contributions, $\mathcal{Q}[\rho _{AB}(t)]$
and $\mathcal{D}[\rho _{AB}(t)].$ In contrast, in Fig 5(b) a similar
behavior to that shown in Fig. 2(d) is recovered. The initial condition is $%
\rho _{AB}(0)=\left\vert ++\right\rangle \left\langle ++\right\vert .$ These
results demonstrate that at short times the relative weights of each
correlation have a strong dependence on the dynamics and initial conditions
under consideration.

\subsection{Breaking symmetries conditions}

While we have considered the projectors defined by Eq. (\ref{projectores})
and (\ref{constraints}), the same result can be achieved with $\mathcal{P}%
_{A}=\mathcal{P}_{B}=\left\vert +\right\rangle \left\langle +\right\vert .\ $%
In this case, the pure stationary state is $\left\vert \Psi
_{+}\right\rangle .$ On the other hand, the achievement of a maximal
entangled state relies on the validity of some symmetry assumptions. Below,
we analyze what happens when they are not meet.

For the dynamics defined by projectors (\ref{constraints}), the assumption $%
\omega _{A}=\omega _{B}$ can be raised up. In fact, by denoting the
frequency of the external (laser) excitation by $\omega _{L},$ instead of
the resonant condition $\omega _{L}=\omega _{A}=\omega _{B},$ for $\omega
_{A}\neq \omega _{B}$ one must to choose $\omega _{L}=(\omega _{A}+\omega
_{B})/2.$ Thus, in an interaction representation with respect to $\hbar
\omega _{L}[\sigma _{z}\otimes \mathrm{I}_{B}+\mathrm{I}_{A}\otimes \sigma
_{z}]/2,$ the Hamiltonian (\ref{HRabi}) acquires the extra contribution $%
\hbar (\omega _{A}-\omega _{B})[\sigma _{z}\otimes \mathrm{I}_{B}-\mathrm{I}%
_{A}\otimes \sigma _{z}]/2.$ Under the condition $\Omega _{A}=\Omega _{B}$
the same maximal entangled state is obtained.%
\begin{figure}[tbp]
\includegraphics[bb=35 150 510
530,angle=0,width=9cm]{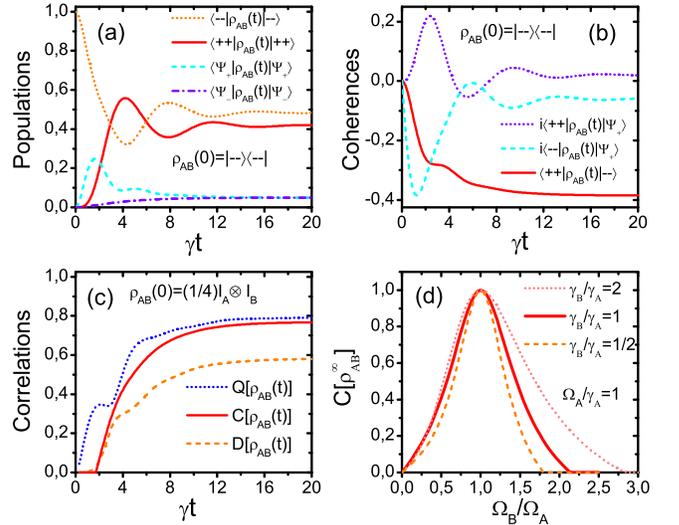}
\caption{(a) Collective populations as function of time. (b) Evolution of
collective coherences. (c) Concurrence and classical-quantum correlations.
(d) Stationary concurrence as a function of $\Omega _{B}/\Omega _{A}$ for
different values of $\protect\gamma _{B}/\protect\gamma _{A}.$ All these
plots correspond to the dynamics (\protect\ref{evolution}) with the
Hamiltonian Eq. (\protect\ref{HRabi}) and the constraints defined by Eqs. (%
\protect\ref{projectores}) and (\protect\ref{constraints}). In plots (a),
(b), and (c) the parameters are $\protect\gamma _{A}=\protect\gamma _{B}=%
\protect\gamma ,$ $\Omega _{A}/\protect\gamma =1,$ and $\Omega _{B}/\protect%
\gamma =3/4.$ The initials conditions \ are indicated in each plot.}
\end{figure}

In Fig. 6 we analyze the stationary entanglement when the Rabi frequencies
are different. We conclude that the previous results are not a singular
property that only happens when $\Omega _{A}=\Omega _{B},$ that is, for $%
|\Omega _{A}-\Omega _{B}|\ll (\Omega _{A}+\Omega _{B})$ it follows $C[\rho
_{AB}^{\infty }]\simeq 1.$

In Fig. 6(a) we show the collective populations. In comparison with Fig.
4(a) (same initials conditions, $\rho _{AB}(0)=\left\vert --\right\rangle
\left\langle --\right\vert ),$ here the population $\left\langle \Psi
_{-}\right\vert \rho _{AB}(t)\left\vert \Psi _{-}\right\rangle $ becomes
populated in the stationary regime, indicating the departure of $\rho
_{AB}^{\infty }$\ from a maximal entangled state. This fact is corroborated
in Fig. 6(b), where $|\left\langle ++\right\vert \rho _{AB}^{\infty
}\left\vert --\right\rangle |<1/2.$ Furthermore, (not shown) coherences
involving the state $\left\vert \Psi _{-}\right\rangle $ also become not
null. These departures arise because the condition $\Omega _{A}\neq \Omega
_{B}$ eliminate the eigenvectors degeneracy of the unitary dynamics [see Eq.
(\ref{Hamilton})].

In Fig. 6(c) we plot the evolution of the concurrence and the
classical-quantum correlations. The initial condition is the identity
matrix, $\rho _{AB}(0)=(\mathrm{I}_{A}\otimes \mathrm{I}_{B})/4.$ As
expected, the concurrence does not converge to one. On the other hand, for
this initial condition a delayed birth of entanglement is observed. As in
previous section, it is preceded by a fast increasing of the classical
correlation and a very small contribution of the quantum correlation.

In Fig. 6(d) we plot the stationary concurrence, $C[\rho _{AB}^{\infty }],$
as a function of $\Omega _{B}/\Omega _{A}$ for different values of $\gamma
_{B}/\gamma _{A}.$ Independently of this last parameter, for $\Omega
_{B}/\Omega _{A}=1$ we get $C[\rho _{AB}^{\infty }]=1.$ On the other hand,
high concurrence values can be achieved around this point. These curves
demonstrate the robustness of the dynamics for generating almost pure
entangled states even when the symmetry conditions are not meet exactly. In
order to quantify this property, we calculated the stationary density matrix
for $\gamma _{A}=\gamma _{B}=\gamma $ and $\Omega _{B}\neq \Omega _{A}.$ To
first order in $\delta \Omega =(\Omega _{A}-\Omega _{B})/2,$ in the product
basis Eq. (\ref{Puros}), we get%
\begin{equation}
\rho _{AB}^{\infty }=\left( 
\begin{array}{cccc}
\frac{1}{2} & -i\frac{\delta \Omega }{\gamma } & i\frac{\delta \Omega }{%
\gamma } & -\frac{1}{2} \\ 
i\frac{\delta \Omega }{\gamma } & 0 & 0 & -i\frac{\delta \Omega }{\gamma }
\\ 
-i\frac{\delta \Omega }{\gamma } & 0 & 0 & i\frac{\delta \Omega }{\gamma }
\\ 
-\frac{1}{2} & i\frac{\delta \Omega }{\gamma } & -i\frac{\delta \Omega }{%
\gamma } & \frac{1}{2}%
\end{array}%
\right) +O(\delta \Omega ^{2}).
\end{equation}%
On the other hand, from the exact expression for $\rho _{AB}^{\infty },$ the
stationary concurrence can be written as%
\begin{equation}
C[\rho _{AB}^{\infty }]=1-\left( \frac{16}{\gamma ^{2}}+\frac{2}{\Omega ^{2}}%
\right) \delta \Omega ^{2}+O[\delta \Omega ^{3}],
\end{equation}%
where $\Omega =(\Omega _{A}+\Omega _{B})/2.$ Therefore, the concurrence
decreases in a quadratic way with the asymmetry $\delta \Omega .$

\section{Summary and Conclusions}

We characterized a class of constrained quantum dissipative evolution where
dynamical constraints are introduced through a set of projectors that
conditioning the action of each dissipative (Lindblad) channel to the state
of the other system, Eqs. (\ref{Lindblad}) and (\ref{Constraint}). When the
set of projectors is closed under the dynamical action of each dissipative
dynamic, Eq. (\ref{closure}), the kinetic constraints can be read in a
classical way, that is, the reduced density matrixes evolve with a Lindblad
rate equation, Eq. (\ref{LindbladRates}), and the conditional expectation
values of the projectors are defined by a classical master equation, Eq. (%
\ref{maestraClasica}).

With the previous ingredients, we studied the stationary entanglement that
can be achieved by two optical-like qubits whose individual decay dynamics
is only possible when the other system is in the lower state. This classical
constraint lead to a free decoherence state, which play a central role in
the entanglement generation. Under the solely action of the dissipative
dynamics, the stationary entanglement depends on the initial condition.
Maximal entangled states are unreachable from separable initial states. On
the contrary, we showed that by coupling the systems to local external
Hamiltonian fields, Eq. (\ref{HRabi}), the interplay between the coherent
and dissipative effects drive the systems to a maximal entangled Bell state,
Eq. (\ref{EstacionBell}). This property does not depends on the system
initialization. The coherent dynamic that couple the free decoherence state
(induced by the constraints) and its complementary space is the central
ingredient that give rise to this property. The underlying mechanism can be
understood in a collective basis, where a Bell state play the role of lower
an upper level for the coherent and incoherent couplings induced by the
external excitation and the constrained dynamics respectively (Fig. 3).

We also studied the time evolution of the entanglement. The present model
confirm that quantum and classical correlations between the systems are
precursors of the entanglement, that is, the arise before entanglement
emerges. Nevertheless, in contrast to other dynamics such as the Dicke model 
\cite{luiz}, here the weights of both correlations may strongly depend on
both the initial conditions and the external excitation.

Our analyses applies to bipartite systems. Due to the present advances in
engineered system-bath coupling constrained dynamics could be checked in
that context. On the other hand, the present results may apply when
restricting the dynamics of a complex many body system to a given bipartite
subspace.




\section*{Acknowledgments}

This work was supported by Consejo Nacional de Investigaciones Cient\'{\i}%
ficas y T\'{e}cnicas (CONICET), Argentina, under Grant No. PIP
11420090100211.

\appendix*

\section{Entanglement measure and quantum-classical correlations}

Here, we provide the definition of the correlation measures used along the
manuscript. The entanglement measure associated to the entanglement of
formation \cite{wooters} is the concurrence. For a bipartite state $\rho
_{AB}$ it is defined as%
\begin{equation}
C[\rho _{AB}]=\max \{0,\sqrt{\lambda _{1}}-\sqrt{\lambda _{2}}-\sqrt{\lambda
_{3}}-\sqrt{\lambda _{4}}\}.  \label{concurrol}
\end{equation}%
$\{\lambda _{1}\}_{i=1}^{i=4}$\ are the eigenvalues (written in decreasing
order) of the matrix $\rho _{AB}\tilde{\rho}_{AB},$ where $\tilde{\rho}%
_{AB}=\sigma _{x}\otimes \sigma _{x}\rho _{AB}^{\ast }\sigma _{x}\otimes
\sigma _{x},$ and $\rho _{AB}^{\ast }$ is the conjugate of $\rho _{AB}$ in
the computational basis. For the stationary state (\ref{rhoInfi}), the
eigenvalues $\{\lambda _{1}\}_{i=1}^{i=4}$ are $\{(\sqrt{p(1-p)}\pm
|c|)^{2},0,0\},$ which leads to Eq. (\ref{ConcuEstacion}).

The mutual information\ provides a measure of the total correlations in a
bipartite setup%
\begin{equation}
\mathcal{I}(\rho _{AB})=S(\rho _{A})+S(\rho _{B})-S(\rho _{AB}),
\end{equation}%
where the Shannon entropy reads $S(\rho )=-\mathrm{Tr}\{\rho \log _{2}\rho
\},$ and $\rho _{A}=\mathrm{Tr}_{B}[\rho _{AB}],$ $\rho _{B}=\mathrm{Tr}%
_{A}[\rho _{AB}].$ $\mathcal{I}(\rho _{AB})$ can be written as the addition
of a classical contribution \cite{vedral}, $\mathcal{Q}(\rho _{AB}),$ and a
quantum contribution or discord \cite{discord}, $\mathcal{D}(\rho _{AB}),$%
\begin{equation}
\mathcal{I}(\rho _{AB})=\mathcal{Q}(\rho _{AB})+\mathcal{D}(\rho _{AB}).
\end{equation}%
The classical correlation is defined as%
\begin{equation}
\mathcal{Q}(\rho _{AB})=\sup_{\{\Pi _{\mathbf{n}}\}}\{S(\rho _{A})-S(\rho
_{AB}|\Pi _{\mathbf{n}})\},  \label{QQ}
\end{equation}%
while the discord reads%
\begin{equation}
\mathcal{D}(\rho _{AB})=\inf_{\{\Pi _{\mathbf{n}}\}}\{\mathcal{I}(\rho
_{AB})-S(\rho _{AB}|\Pi _{\mathbf{n}})\}.  \label{DD}
\end{equation}%
In both expressions, $S(\rho _{AB}|\Pi _{\mathbf{n}})$ refers to the
conditional entropy of system $A$ given that a measurement over system $B$
was performed. This action can be defined by a measurement in an arbitrary
direction $\mathbf{n}$ over the Bloch sphere. Thus, $\{\Pi _{\mathbf{n}%
}\}=\Pi _{\mathbf{n}\pm }(\theta ,\phi )=\left\vert \mathbf{n}_{\pm
}\right\rangle \left\langle \mathbf{n}_{\pm }\right\vert ,$ where 
\begin{subequations}
\label{eigenbloch}
\begin{eqnarray}
\left\vert \mathbf{n}_{+}\right\rangle &=&+\cos (\theta /2)\left\vert
+\right\rangle +e^{+i\phi }\sin (\theta /2)\left\vert -\right\rangle ,\ \ \ 
\\
\left\vert \mathbf{n}_{-}\right\rangle &=&-\sin (\theta /2)\left\vert
+\right\rangle +e^{i\phi }\cos (\theta /2)\left\vert -\right\rangle .
\end{eqnarray}%
In this way, the extremization of Eqs. (\ref{QQ}) and (\ref{DD}) is
performed over the polar angles $\theta $ and $\phi .$

For the stationary state $\rho _{AB}^{\infty },$ Eq. (\ref{rhoInfi}), it is
simple to realize that it is always possible to chose a projector $\Pi _{%
\mathbf{n}}$ $(\theta =0,\pi ,\pi /2)$ such that the minimal value of the
conditional entropy $S(\rho _{AB}|\Pi _{\mathbf{n}})$\ is cero. Hence, it
follows 
\end{subequations}
\begin{equation}
\mathcal{Q}(\rho _{AB}^{\infty })=S(\rho _{A}^{\infty })=S(\rho _{B}^{\infty
}),  \label{ClasicaEst}
\end{equation}%
while the stationary discord becomes%
\begin{equation}
\mathcal{D}(\rho _{AB}^{\infty })=S(\rho _{A}^{\infty })-S(\rho
_{AB}^{\infty })=S(\rho _{B}^{\infty })-S(\rho _{AB}^{\infty }).
\label{DiscordEst}
\end{equation}%
The entropies $S(\rho )=-\mathrm{Tr}\{\rho \log _{2}\rho \}$ can trivially
be determine from the eigenvalues of the corresponding matrixes. The four
eigenvalues of $\rho _{AB}^{\infty }$ are $\{\frac{1}{2}[1\pm \sqrt{%
1-4p(1-p)+4|c|^{2}}],0,0\}.$ From here, one deduce that the stationary
discord $\mathcal{D}(\rho _{AB}^{\infty })$ is not null only when $|c|\neq
0, $ that is, the same condition for getting a non null stationary
concurrence, Eq. (\ref{ConcuEstacion}).

The initial conditions and dynamics considered along the paper lead to a
bipartite density matrix\ $\rho _{AB}$ with the structure 
\begin{equation}
\rho _{AB}=\left( 
\begin{array}{cccc}
q & -id & -id^{\prime } & w \\ 
id & r & u & -iv \\ 
id^{\prime } & u & r^{\prime } & -iv^{\prime } \\ 
w & iv & iv^{\prime } & s%
\end{array}%
\right) ,
\end{equation}%
where each letter represent a real number. For symmetric states, $\rho
_{A}=\rho _{B},$ it follows $d^{\prime }=d,$ $v^{\prime }=v,$ $r^{\prime
}=r. $ The concurrence of this kind of state can be calculated in an
analytical way. Nevertheless, the expression becomes very complicated and
does not provide useful information. On the other hand, we find that $S(\rho
_{AB}|\Pi _{\mathbf{n}})$ is minimal (with the same value) at angles $\phi
=\pi /2$ and $\phi =3\pi /2.$ Nevertheless, consistently with the analysis
of Ref. \cite{adesso} the value of $\theta $ must to be found in a numerical
way. When $\rho _{AB}$ adopts a X structure, $\{d=d^{\prime }=v=v^{\prime
}=0\},$ the implemented numerical algorithm recovers the analytical discord
values presented in Ref. \cite{caldeira}.

\end{document}